\documentclass[prl,reprint,superscriptaddress]{revtex4-1}

\usepackage{amsmath,amssymb}
\usepackage{verbatim}
\usepackage{graphicx}
\usepackage{hyperref}
\usepackage{color}
\usepackage{mathrsfs}

\newcommand {\cL}{{\cal L}}

\def\a{\alpha}

\def\d{\delta}

\def\f{\phi}

\def\l{\lambda}
\def\m{\mu}
\def\n{\nu}

\def\r{\rho}
\def\s{\sigma}
\def\t{\tau}

\def\P{\Pi}

\def\oneone{\rlap 1\mkern4mu{\rm l}}

\newcommand{\be}{\begin{equation}}
	\newcommand{\ee}{\end{equation}}
\newcommand{\bea}{\begin{eqnarray}}
	\newcommand{\eea}{\end{eqnarray}}

\newcommand{\ba}{\begin{array}}
	\newcommand{\ea}{\end{array}}

\def\double #1{#1{\hbox{\kern-2pt $#1$}}}

\newcommand{\bsubeq}{\begin{subequations}}
	\newcommand{\esubeq}{\end{subequations}}

\def\ft#1#2{{\textstyle{\frac{\scriptstyle #1}{\scriptstyle #2} } }}

\def\rmi{{\rm i}}

\def\a{\alpha}

\def\d{\delta}

\def\f{\phi}

\def\P{\Psi}

\def\l{\lambda}

\def\m{\mu}
\def\n{\nu}
\def\r{\rho}
\def\s{\sigma}

\def\t{\tau}

\begin{document}
	
	\title{Carrollian Origin of Spacetime Subsystem Symmetry}
	
	\author{Oguzhan Kasikci}
	\email{kasikcio@itu.edu.tr}
	\affiliation{Department of Physics,
		Istanbul Technical University,
		Maslak 34469 Istanbul,
		Turkey}
	
	\author{Mehmet Ozkan}
	\email{ozkanmehm@itu.edu.tr}
	\affiliation{Department of Physics,
		Istanbul Technical University,
		Maslak 34469 Istanbul,
		Turkey}
	
	\author{Yi Pang}
	\email{pangyi1@tju.edu.cn}
	\affiliation{Center for Joint Quantum Studies and Department of Physics,\\
		School of Science, Tianjin University, Tianjin 300350, China \\}

	\date{\today}
	

	\begin{abstract}
		
		We propose that models with spacetime subsystem symmetry are connected to  Lorentz invariant models via the Carrollian limit. In this way, a recently proposed model with spacetime subsystem symmetry was readily reproduced together with its conserved charges.  We then couple this model to a dynamical Abelian gauge field and  Carroll gravity.  Our procedure can be applied in arbitrary dimensions and paves the way to construct new models with spacetime subsystem symmetry.
		
	\end{abstract}
	
	
	\maketitle
	\allowdisplaybreaks
	
	In recent years, it has become evident that the Carroll symmetry provides a new set of tools to explore various phenomena that were previously not considered from a symmetry viewpoint. Initially introduced as the Poincar\'e group in the limit of vanishing speed of light $(c \to 0)$ \cite{LLJM1,LLJM2}, the characteristic feature of Carrollian physics is that a Carroll particle cannot move. This ultra-local behavior  has been proposed to explain the vanishing of the Love numbers for the Schwarzschild black hole \cite{Penna:2018gfx},  the inflationary regime in the early universe \cite{deBoer:2021jej} and fractons \cite{Bidussi:2021nmp,Marsot:2022imf}. The notion of Carroll geometry can also be introduced in a similar manner by taking the $c\to 0$ limit of a Lorentzian metric \cite{Dautcourt:1997hb} which has been shown to arise in physics of null hypersurfaces \cite{Duval:2014uoa,Hartong:2015xda,Blau:2015nee,Ciambelli:2019lap} utilized to study the dynamics of black hole horizon \cite{Donnay:2019jiz}.  Furthermore, when extended with the conformal generators, the Carroll group is known to be isomorphic to the Bondi-Metzner-Sachs (BMS) group \cite{Duval:2014uva} which plays a central role in flat space holography \cite{Bagchi:2013bga,Bagchi:2015nca,Bagchi:2016bcd,Saha:2023hsl} and in the IR behavior of gravity \cite{Strominger:2017zoo,Fuentealba:2022gdx}. 
	
	In this paper, we aim to connect Carroll symmetry to another corner of physics. We show that continuum field theory models exhibiting the spacetime subsystem symmetry are Carrollian. The notion of spacetime subsystem symmetry is a generalization of the internal subsystem symmetry initially discovered in lattice models characterizing exotic phases of matter. By definition, subsystem symmetries are generated by charges conserved on lines, planes, or other substructures. Inspired by  models with internal subsystem symmetries, the very recent work \cite{Baig:2023yaz} proposed a 2+1 dimensional field theory model
	\cite{Baig:2023yaz}
	\be
	{\cal L}=\ft12\dot{\phi}_1^2+\ft12\dot{\phi}_2^2+\ft12(\dot{\phi_1}\partial_i\phi_2- \dot{\phi_2}\partial_i\phi_1  )^2-V(\phi_1,\phi_2)\ ,
	\label{MainLag}
	\ee
	equipped with the spacetime subsystem symmetry 
	\be
	t\rightarrow t+c(x,y),\quad x,\,y\rightarrow x,\,y\ .
	\label{trans}
	\ee
	The symmetry transformation above implies that the model admits a locally conserved Hamiltonian. In this paper, we show that the spacetime subsystem symmetry is in fact associated with the so-called Carroll boost, and the Lagrangian \eqref{MainLag} can be obtained in the vanishing speed of light limit $(c \to 0)$ of a Poincar\'e invariant theory.  
	
	\textit{An Observation---} 	The distinctive feature of the Carrollian symmetry is that aside from the standard time translations and spatial rotations, it contains Carrollian boosts that originate from the Lorentz boosts in the $c\to 0$ limit, given by $C_i = x_i \partial_t$ where $i = 1, 2$ represent the spatial dimensions \cite{Duval:2014uoa}. The structure of the Carrollian boosts implies that while the spatial coordinates are inert, the time coordinate transforms as $t \to t + \vec c \cdot\vec x$ whose local form coincides with the spacetime subsystem symmetry \eqref{trans}.	This fact suggests that the model \eqref{MainLag} may arise from the ultra-relativistic limit $(c \to 0)$ of a relativistic model. To show that this is indeed the case, we first restore the $c$ dependence of the $d+1$ dimensional Minkowski metric 
	\begin{align}
		\eta_{\m\n} &= 	\begin{pmatrix}
			- c^2 & 0\\
			0 & \oneone_{d\times d}
		\end{pmatrix}\,, & 	\eta^{\m\n} &= 	\begin{pmatrix}
			- \frac{1}{c^2} & 0\\
			0 & \oneone_{d\times d}
		\end{pmatrix}\,. & 
	\end{align}
	Starting from the following Lagrangian in $(d+1)$-dimensions 
	\begin{eqnarray}
		\cL = - \frac12 \partial_\m \f \partial^\m \f  =  \frac1{2 c^2} \dot\phi^2 - \frac12 \partial_i \phi \partial_i \phi \,,
		\label{2derivative}
	\end{eqnarray}
	after redefining the  scalar fields $\f_1 \to c \f_1$ and $\f_2 \to c \f_2$ and taking the limit $c \to 0$, we obtain the  Carroll invariant quadratic action
	\begin{eqnarray}
		\cL_{2\partial} &=& \frac12 \dot\phi^2 \,,
	\end{eqnarray}
	which has also been obtained from other means \cite{Bidussi:2021nmp,Chen:2023pqf}.
	Thus, the ultra-relativistic limit of the Lagrangian
	\begin{eqnarray}
		\cL = - \frac12 \partial_\m \f_1 \partial^\m \f_1 - \frac12 \partial_\m \f_2 \partial^\m \f_2 \,,
	\end{eqnarray} 
	give rise to the two-derivative terms in \eqref{MainLag}. Next, we turn to the four-derivative part in \eqref{MainLag}. Consider the following combination
	\begin{eqnarray}
		\cL_{4\partial} &=& \frac{\alpha}2 (\partial_\m \f_1 \partial^\m \f_2)^2 -  \frac{\alpha}2 (\partial_\m \f_1 \partial^\m \f_1) 	(\partial_\n \f_2 \partial^\n \f_2) \nonumber\\
		\nonumber\\
		&=&  \frac{\alpha}{2c^2} \big( \dot\f_1 \partial_i \f_2 - \dot\f_2 \partial_i \f_1 \big)^2 + \mathcal{O}(c^0) \,,
	\end{eqnarray}
	where $\a$ is a positive constant of dimension $[{\rm length}]^{d+1}$.
	Consequently, rescaling the scalar fields $\f_1 \to c \f_1\,, \f_2 \to c \f_2$ and the parameter $\alpha \to \alpha/c^2$ followed by the ultra-relativistic limit leads precisely to the four-derivative terms in  \cite{Baig:2023yaz}. Thus, the ultra-relativistic limit of the Lagrangian below
	\begin{eqnarray}
		\cL &=& - \frac12 \partial_\m \f_1 \partial^\m \f_1 - \frac12 \partial_\m \f_2 \partial^\m \f_2 + \frac{\a}2 (\partial_\m \f_1 \partial^\m \f_2)^2 \nonumber\\
		&&  -  \frac{\a}2 (\partial_\m \f_1 \partial^\m \f_1) 	(\partial_\n \f_2 \partial^\n \f_2) \,, 
		\label{Karch}
	\end{eqnarray}
	gives rise to the kinetic terms of the model given in \cite{Baig:2023yaz}.
	Assuming $V(\phi_i)$ is a polynomial in $\phi_i$, then 
	under the rescaling of scalar fields $\phi_i\rightarrow c\phi_i$ and the proper rescaling of the coupling constants in $V(\phi_i)$, a nonvanishing potential term can be acquired. From now on, we will omit the potential term, which can be easily added back without affecting the symmetry properties of the model. 
	
	\textit{Conserved currents---} 
	The Noether charges of the relativitic model can be easily computed by varying respect to the background metric.  For the Lagrangian \eqref{Karch}, the energy-momentum tensor is given by
	\begin{eqnarray}
		T^\m{}_\n &=& - \partial^\m \f_1 \partial_\n \f_1  - \partial^\m \f_2 \partial_\n \f_2- \d^\m{}_\n \cL  \nonumber\\
		&& + \a \big( - \partial^\m \f_1 \partial_\n \f_1 \partial^\l \f_2 \partial_\l \f_2  - \partial^\m \f_2 \partial_\n \f_2 \partial^\l \f_1 \partial_\l \f_1 \nonumber\\
		&& + \partial^\m \f_1 \partial_\n \f_2 \partial^\l \f_1 \partial_\l \f_2  +  \partial^\m \f_2 \partial_\n \f_1 \partial^\l \f_1 \partial_\l \f_2 \big)  \,.
	\end{eqnarray}
	To obtain the conserved currents of the ultra-relativistic model, we perform the the rescaling of $\phi$ and $\a$ stated before. After this is done,  we find that the Hamiltonian density is of the form
	\begin{eqnarray}
		T^0{}_0 &=& \frac12 \dot\f_1^2 + \frac12 \dot\f_2^2 + \frac12 \a \chi_i^2+{\cal O}(c^2) \,.
		\label{T00}
	\end{eqnarray}
	where $\chi_i$ is 
	\begin{eqnarray}
		\chi_i = (\dot\f_2 \partial_i \f_1 - \dot \f_1 \partial_i \f_2 ) \,.
	\end{eqnarray}
	Hence its ultra-relativistic limit reproduces the Hamiltonian density found by \cite{Baig:2023yaz}. Next, we have
	\begin{eqnarray}
		T^i{}_0 &=& - c^2 ( \dot \f_1 \partial^i \f_1 + \dot\f_2 \partial^i \f_2 ) \nonumber\\
		&& + c^2 \alpha ( \chi_j \partial_i \f_2 \partial_j \f_1-\chi_j \partial_i \f_1 \partial_j \f_2 )\,, 
	\end{eqnarray}
	Thus this component vanish in the $c \to 0$ limit, i.e. $T^i{}_0 = 0$ for the model \eqref{Karch} meaning the the energy does not flow in this model. The momentum density $T^{0}{}_i$ is given by
	\begin{eqnarray}
		T^0{}_i &=& - \frac{1}{c^2} T_{0}{}^i \,,
	\end{eqnarray}
	which upon taking the limit $c\rightarrow 0$ is identical to the one given in \cite{Baig:2023yaz}. Finally, for $T^i{}_j$, the $c\to 0$ limit is given by
	\begin{eqnarray}
		T^i{}_j &=&  \alpha \chi^i \chi_j  - \d^{i}{}_j \big(\frac12 \dot\f_1^2 + \frac12 \dot \f_2^2 + \frac12 \a\chi_k^2\big) \,.
	\end{eqnarray}
	which differs from the the result given in \cite{Baig:2023yaz} by a term proportional to $\delta^i_j$. One can check that for the momentum current to be conserved, this term is indispensable.  
	
	\textit{Coupling to a U(1) gauge field---} We may rewrite the ultra-relativistic Lagrangian \eqref{MainLag} by defining a complex scalar field 
	\begin{eqnarray}
		\P = \frac{1}{\sqrt2} \left(\f_1 + \rmi \f_2 \right) 
	\end{eqnarray}
	in terms of which we have
	\begin{eqnarray}
		\cL &=& \partial_t \P \, \partial_t \bar\P - \frac{\a}2 \left(\partial_t \P \, \partial_i \bar\P - \partial_t \bar\P \, \partial_i \P \right)^2 \,.
	\end{eqnarray}
	Note that this model is invariant under Carrollian boosts but not under the $U(1)$-dipole symmetry 
	\begin{eqnarray}
		\P \to e^{\rmi \vec{b}\cdot \vec{x}} \P\,. 
	\end{eqnarray}
	The relativistic origin of the complex scalar model is given by
		\begin{eqnarray}
			\cL &=& - \partial_\mu \P \partial^\m \bar \P - \frac{\alpha}{2} \partial_\m \P \partial^\m \bar\P  \partial_\n \P \partial^\n \bar\P \nonumber\\
			&& + \frac{\alpha}{2} \partial_\m \P \partial^\m \P  \partial_\n \bar\P \partial^\n \bar\P  \,.
		\end{eqnarray}
		For this model, one can add a U(1) invariant potential term. 
		If we gauge this model, we have
			\begin{eqnarray}
				\cL &=& - D_\mu \P D^\m \bar \P - \frac{\alpha}{2} D_\m \P D^\m \bar\P  D_\n \P D^\n \bar\P \nonumber\\
				&& + \frac{\alpha}{2} D_\m \P D^\m \P  D_\n \bar\P D^\n \bar\P -  \frac14 F_{\m\n} F^{\m\n} \,.
			\end{eqnarray}
			Here, $F_{\m\n} = \partial_\m A_\n - \partial_\n A_\m$ and $D_\m \P= \partial_\m \P - \rmi g A_\m \P$.  To perform the electric limit of the theory, we rescale the fields $\P \to c \P, A_\m \to c A_\m$ and redefine $\alpha \to \alpha/c^2$ and $g \to g/c$ in which case the $c \to 0$ limit gives rise to
			\begin{eqnarray}
				\cL_{U(1)} &=&  D_t \P D_t \bar\P - \frac12 E_i^2   \nonumber\\
				&& - \frac{\a}2 \left(D_t \P \, D_i  \bar\P - D_t  \bar\P \, D_i \P \right)^2 
			\end{eqnarray}
			where we defined
			\begin{align}
				D_t \P &= \partial_t \P -  \rmi g A_t \P   \,, &  D_i \P  &= \partial_i \P -  \rmi g A_i \P \,,\nonumber\\
				E_i & = \partial_t A_i - \partial_i A_t \,,
			\end{align}
			where the action for the U(1) gauge field also appeared in previous works  \cite{Duval:2014uoa, Henneaux:2021yzg} as the electric-like contraction of Maxwell theory.

			\textit{Coupling to Carrollian gravity--- } Carroll gravity is the ultra-relativistic limit of General Relativity \cite{Dautcourt:1997hb}. The fundamental fields of the theory are given by an inverse temporal vielbein $\tau^\mu$, a spatial vielbein $e_\m{}^a\,,$ with $a=1,2$ and $\mu = 0,1,2$ \cite{Henneaux:1979vn} and a Lagrange multiplier that imposes vanishing of the spatial part of the extrinsic curvature \cite{Bergshoeff:2017btm,Campoleoni:2022ebj}. The spatial and temporal vielbein, along with their inverses satisfy the following relations
			\begin{align}
				e_\m{}^a e^\m{}_b & = \d^{a}_{b} \,, & \t^\m \t_\m & = 1\,, \qquad~ \t^\m e_\m{}^a = 0 \,,
				\nonumber\\
				\t_\m e^\m{}_a & = 0 \,, & e_\m{}^a e^\n{}_a & = \d_\m{}^\n - \tau_\m \tau^\n \,,
				\label{etaurelations}
			\end{align}
			and the non-vanishing Carroll boost transformations are
			\begin{align}
				\d \t_\m & = \l_a e_\m{}^a\,, & \d e^\m{}_a & = - \l_a \t^\m \,, 
			\end{align}
			Utilizing the orthogonality relations \eqref{etaurelations}, we can decompose a vector $V_\m$ using the temporal and spatial projections according to 
			\begin{eqnarray}
				V_\m = \tau_\m V_t + e_\m{}^a V_a \,.
			\end{eqnarray}
			These relations together with the fact that the scalar fields $\phi_{1,2}$ are inert under local Carrollian boosts imply the temporal and spatial derivatives transform as
			\begin{eqnarray}
				\delta \left(\tau^\m \partial_\m \phi \right) = 0 \,, \qquad \delta \left( e^\m{}_a \partial_\m \phi\right) = - \lambda_a \tau^\m \partial_\m \phi \,.
			\end{eqnarray}
			Consequently, we may give the covariant formulation of the subsystem invariant action as
			\begin{eqnarray}
				e^{-1}	\cL &=& \frac12 \tau^\m \tau^\n \left( \partial_\m \f_1 \partial_\n \f_1  + \partial_\m \f_2 \partial_\n \f_2 \right) \nonumber\\
				&& + \frac12 \big(2 \tau^{[\m }e^{\n]}{}_a  \partial_\m \f_1 \partial_\n \f_2\big)^2 \,,
			\end{eqnarray}
			where $e = \det (\t_\m \t_\n + e_\m{}^a e_{\n a})$ is ultra-relativistic determinant.  The variation of the action with respect to $\tau^\m$ gives rise to
			\begin{eqnarray}
				T_\m &=& \t^\n \left( \partial_\m \f_1 \partial_\n \f_1 + \partial_\m \f_2 \partial_\n \f_2\right) \nonumber\\
				&& + \t^\rho h^{\n\s} \chi_{\m\n} \chi_{\r\s} - \tau_\m \left(e^{-1}\cL\right) \,,
			\end{eqnarray}
			where we defined $h^{\m\n} = e^{\m}{}_a e^{\n a}$ and
			\begin{eqnarray}
				\chi_{\m\n} = \partial_\m \f_1 \partial_\n \f_2 - \partial_\n \f_1 \partial_\m \f_2 \,. 
			\end{eqnarray}
			The temporal component of $T_\m$ is then given by
			\begin{eqnarray}
				T_0 &=& \tau^\m T_\m \nonumber\\
				&=& \frac12  \tau^\m \tau^\n \left( \partial_\m \f_1 \partial_\n \f_1  + \partial_\m \f_2 \partial_\n \f_2 \right)  \nonumber\\
				&& + \frac12 \t^\m \t^\rho h^{\n\s} \chi_{\m\n} \chi_{\r\s}\, .
			\end{eqnarray}
			Furthermore, the spatial component of $T_\m$ gives rise to the conserved momentum
			\begin{eqnarray}
				T_b &=& e^\m{}_b T_\m\nonumber\\
				&=& e^\m{}_b\t^\n \left( \partial_\m \f_1 \partial_\n \f_1 + \partial_\m \f_2 \partial_\n \f_2\right) \nonumber\\
				&& + e^\m{}_b \t^\rho h^{\n\s} \chi_{\m\n} \chi_{\r\s} 
			\end{eqnarray}
			Next, we turn to the variation with respect to $e^\m{}_a$. As this field is independent of $\tau^\m$, the variation is unconstrained and is given by
			\begin{eqnarray}
				T_\m{}^a &=& - e_\m{}^a \cL + e^{\n a} \tau^\s \t^\r \chi_{\s\n} \chi_{\r\m}
			\end{eqnarray}
			which gives rise to the following components
			\begin{eqnarray}
				T_{0}{}^a &=& \tau^\m T_{\m}{}^a = 0 \,,\nonumber\\
				T_{b}{}^a &=& e^\m{}_b T_{\m}{}^a = - \delta_b{}^a \cL + e^\m{}_b e^{\n a} \t^\r \t^\s \chi_{\s\n} \chi_{\r\m} \,.
			\end{eqnarray}
			These results precisely match with the ones given in components in the previous section upon substituting $\tau^\m=\delta^\m_0,\, e^{\m}{}_b=\delta^\m_b$.

			\textit{Discussion and outlook}--- In this paper, we show that continuum field theory models with spacetime subsystem symmetry can arise from an appropriate Carrollian limit of Poincare invariant models, facilitating the construction of such models dramatically. Once the field theory models are known, the corresponding lattice models can be reconstructed. We demonstrate our method by obtaining the recently proposed model with spacetime subsystem symmetry by properly taking the $c\rightarrow 0$ limit of a relativistic two-scalar model. We coupled the model to an Abelian gauge field when the scalar potential preserves the U(1) symmetry. Coupling to Carroll gravity is also achieved for this model.

			The model we studied here consists of two scalar fields. One may wonder if  a single field interacting model with spacetime subsystem symmetry exists.  In fact, the other model proposed in \cite{Baig:2023yaz} with the following 4-derivative part
			\begin{eqnarray}
				\cL =  \f \left[ \left(\partial_t^2 \f \right) \left( \partial^i \partial_i \f \right) - \left( \partial_i \partial_t \f \right) \left( \partial^i \partial^t \f \right) \right] \,,
				\label{singefield}
			\end{eqnarray}
			can be obtained from the Carroll limit of the following relativistic model
			\begin{eqnarray}
				\cL &=& \f \left[ \Box \f \Box \f - \left(\nabla_\m \nabla_\n \f \right) \left( \nabla^\m \nabla^\n \f \right) \right] \,.
				\label{model2}
			\end{eqnarray}
			Its couplings to $U(1)$ gauge field and the Carroll gravity can also be achieved parallel to our discussions in the previous sections. Thus, our statement on the connection between Carrollian physics and subsystem symmetries seem to hold in general.  
			
			The single field \eqref{singefield} and the two-field \eqref{Karch} models are in some sense the simplest structure beyond the 2-derivative level with manifest spacetime subsystem symmetry as they contains only the temporal and spatial vielbeins without their derivatives. Since \eqref{Karch} and \eqref{model2} are closely related to Horndeski gravity, it is conceivable that the $c \to 0$ limit of a large class of Horndeski gravity \cite{Horndeski:1974wa} may yield a more complicated single field model with spacetime subsystem symmetry. It is also worthwhile to mention that the relativistic origin of the single-field model \eqref{model2} is not exactly a Horndeski theory although its ultra-relativistic counterpart does have  equations of motion second-order in time derivatives. This is due to the fact that the higher-order time derivatives cancel out due to relative sign between two terms in the Lagrangian and the higher-order spatial derivatives vanish in the $c \to 0$ limit. Thus, it is worthwhile to study Carrollian Horndeski models from a purely ultra-relativistic perspective rather than performing the $c\to 0$ limit.
			
			We would also like to consider models with both spacetime subsystem symmetry and conformal symmetry which may arise from the ultra-relativistic limit of certain relativistic conformal field theories. 
			Of course, the most interesting is the quantization of models with spacetime subsystem symmetry. At first sight, one cannot directly apply the standard perturbative approach since the quadratic term of the model \eqref{MainLag} contains only $\dot{\phi}$ and a  propagator in spacetime is lacking. The fact that the Hamiltonian is positive definite suggests a well-defined quantum theory exists and knowing the relativistic origin of the model should be helpful in its quantization. We leave this interesting for future exploration.

			\textit{Acknowledgements.}--- We are grateful to Utku Zorba for feedback and comments on the manuscript. M.O. is supported in part by TUBITAK grant 121F064 and Istanbul Technical University Research Fund under grant number TGA-2020-42570. M.O. acknowledges the support by the Distinguished Young Scientist Award BAGEP of the Science Academy. M.O. also acknowledges the support by the Outstanding Young Scientist Award of the Turkish Academy of Sciences (TUBA-GEBIP). The work of Y.P. is supported in part by National Natural Science Foundation of China (NSFC) under grant No. 12175164. Y.P. also acknowledges support by the National Key Research and Development Program under grant No. 2022YFE0134300.

			\bibliographystyle{utphys}
			\bibliography{ref}

\providecommand{\href}[2]{#2}\begingroup\raggedright\begin{thebibliography}{10}

\bibitem{LLJM1}
J.-M. L\'evy-Leblond, ``{Une nouvelle limite non-relativiste du groupe de
  Poincaré},'' {\em Ann. Inst. Henri Poincaré III} {\bfseries 3} (1965) 1.

\bibitem{LLJM2}
J.-M. L\'evy-Leblond, ``{Possible kinematics},'' {\em J. Math. Phys.}
  {\bfseries 9} (1968) 1605.

\bibitem{Penna:2018gfx}
R.~F. Penna, ``{Near-horizon Carroll symmetry and black hole Love numbers},''
  \href{http://arxiv.org/abs/1812.05643}{{\ttfamily arXiv:1812.05643
  [hep-th]}}.

\bibitem{deBoer:2021jej}
J.~de~Boer, J.~Hartong, N.~A. Obers, W.~Sybesma, and S.~Vandoren, ``{Carroll
  Symmetry, Dark Energy and Inflation},''
  \href{http://dx.doi.org/10.3389/fphy.2022.810405}{{\em Front. in Phys.}
  {\bfseries 10} (2022) 810405},
  \href{http://arxiv.org/abs/2110.02319}{{\ttfamily arXiv:2110.02319
  [hep-th]}}.

\bibitem{Bidussi:2021nmp}
L.~Bidussi, J.~Hartong, E.~Have, J.~Musaeus, and S.~Prohazka, ``{Fractons,
  dipole symmetries and curved spacetime},''
  \href{http://dx.doi.org/10.21468/SciPostPhys.12.6.205}{{\em SciPost Phys.}
  {\bfseries 12} no.~6, (2022) 205},
  \href{http://arxiv.org/abs/2111.03668}{{\ttfamily arXiv:2111.03668
  [hep-th]}}.

\bibitem{Marsot:2022imf}
L.~Marsot, P.~M. Zhang, M.~Chernodub, and P.~A. Horvathy, ``{Hall effects in
  Carroll dynamics},'' \href{http://arxiv.org/abs/2212.02360}{{\ttfamily
  arXiv:2212.02360 [hep-th]}}.

\bibitem{Dautcourt:1997hb}
G.~Dautcourt, ``{On the ultrarelativistic limit of general relativity},'' {\em
  Acta Phys. Polon. B} {\bfseries 29} (1998) 1047--1055,
  \href{http://arxiv.org/abs/gr-qc/9801093}{{\ttfamily arXiv:gr-qc/9801093}}.

\bibitem{Duval:2014uoa}
C.~Duval, G.~W. Gibbons, P.~A. Horvathy, and P.~M. Zhang, ``{Carroll versus
  Newton and Galilei: two dual non-Einsteinian concepts of time},''
  \href{http://dx.doi.org/10.1088/0264-9381/31/8/085016}{{\em Class. Quant.
  Grav.} {\bfseries 31} (2014) 085016},
  \href{http://arxiv.org/abs/1402.0657}{{\ttfamily arXiv:1402.0657 [gr-qc]}}.

\bibitem{Hartong:2015xda}
J.~Hartong, ``{Gauging the Carroll Algebra and Ultra-Relativistic Gravity},''
  \href{http://dx.doi.org/10.1007/JHEP08(2015)069}{{\em JHEP} {\bfseries 08}
  (2015) 069}, \href{http://arxiv.org/abs/1505.05011}{{\ttfamily
  arXiv:1505.05011 [hep-th]}}.

\bibitem{Blau:2015nee}
M.~Blau and M.~O'Loughlin, ``{Horizon Shells and BMS-like Soldering
  Transformations},'' \href{http://dx.doi.org/10.1007/JHEP03(2016)029}{{\em
  JHEP} {\bfseries 03} (2016) 029},
  \href{http://arxiv.org/abs/1512.02858}{{\ttfamily arXiv:1512.02858
  [hep-th]}}.

\bibitem{Ciambelli:2019lap}
L.~Ciambelli, R.~G. Leigh, C.~Marteau, and P.~M. Petropoulos, ``{Carroll
  Structures, Null Geometry and Conformal Isometries},''
  \href{http://dx.doi.org/10.1103/PhysRevD.100.046010}{{\em Phys. Rev. D}
  {\bfseries 100} no.~4, (2019) 046010},
  \href{http://arxiv.org/abs/1905.02221}{{\ttfamily arXiv:1905.02221
  [hep-th]}}.

\bibitem{Donnay:2019jiz}
L.~Donnay and C.~Marteau, ``{Carrollian Physics at the Black Hole Horizon},''
  \href{http://dx.doi.org/10.1088/1361-6382/ab2fd5}{{\em Class. Quant. Grav.}
  {\bfseries 36} no.~16, (2019) 165002},
  \href{http://arxiv.org/abs/1903.09654}{{\ttfamily arXiv:1903.09654
  [hep-th]}}.

\bibitem{Duval:2014uva}
C.~Duval, G.~W. Gibbons, and P.~A. Horvathy, ``{Conformal Carroll groups and
  BMS symmetry},'' \href{http://dx.doi.org/10.1088/0264-9381/31/9/092001}{{\em
  Class. Quant. Grav.} {\bfseries 31} (2014) 092001},
  \href{http://arxiv.org/abs/1402.5894}{{\ttfamily arXiv:1402.5894 [gr-qc]}}.

\bibitem{Bagchi:2013bga}
A.~Bagchi, ``{Tensionless Strings and Galilean Conformal Algebra},''
  \href{http://dx.doi.org/10.1007/JHEP05(2013)141}{{\em JHEP} {\bfseries 05}
  (2013) 141}, \href{http://arxiv.org/abs/1303.0291}{{\ttfamily arXiv:1303.0291
  [hep-th]}}.

\bibitem{Bagchi:2015nca}
A.~Bagchi, S.~Chakrabortty, and P.~Parekh, ``{Tensionless Strings from
  Worldsheet Symmetries},''
  \href{http://dx.doi.org/10.1007/JHEP01(2016)158}{{\em JHEP} {\bfseries 01}
  (2016) 158}, \href{http://arxiv.org/abs/1507.04361}{{\ttfamily
  arXiv:1507.04361 [hep-th]}}.

\bibitem{Bagchi:2016bcd}
A.~Bagchi, R.~Basu, A.~Kakkar, and A.~Mehra, ``{Flat Holography: Aspects of the
  dual field theory},'' \href{http://dx.doi.org/10.1007/JHEP12(2016)147}{{\em
  JHEP} {\bfseries 12} (2016) 147},
  \href{http://arxiv.org/abs/1609.06203}{{\ttfamily arXiv:1609.06203
  [hep-th]}}.

\bibitem{Saha:2023hsl}
A.~Saha, ``{Carrollian Approach to $1+3$D Flat Holography},''
  \href{http://arxiv.org/abs/2304.02696}{{\ttfamily arXiv:2304.02696
  [hep-th]}}.

\bibitem{Strominger:2017zoo}
A.~Strominger, ``{Lectures on the Infrared Structure of Gravity and Gauge
  Theory},'' \href{http://arxiv.org/abs/1703.05448}{{\ttfamily arXiv:1703.05448
  [hep-th]}}.

\bibitem{Fuentealba:2022gdx}
O.~Fuentealba, M.~Henneaux, P.~Salgado-Rebolledo, and J.~Salzer, ``{Asymptotic
  structure of Carrollian limits of Einstein-Yang-Mills theory in four
  spacetime dimensions},''
  \href{http://dx.doi.org/10.1103/PhysRevD.106.104047}{{\em Phys. Rev. D}
  {\bfseries 106} no.~10, (2022) 104047},
  \href{http://arxiv.org/abs/2207.11359}{{\ttfamily arXiv:2207.11359
  [hep-th]}}.

\bibitem{Baig:2023yaz}
S.~A. Baig, J.~Distler, A.~Karch, A.~Raz, and H.-Y. Sun, ``{Spacetime Subsystem
  Symmetries},'' \href{http://arxiv.org/abs/2303.15590}{{\ttfamily
  arXiv:2303.15590 [hep-th]}}.

\bibitem{Note1}
The original model has a minus sign in front of the $\phi ^4$ term.
  Consequently, the Hamiltonian is not positive definite.

\bibitem{Chen:2023pqf}
B.~Chen, R.~Liu, H.~Sun, and Y.-f. Zheng, ``{Constructing Carrollian Field
  Theories from Null Reduction},''
  \href{http://arxiv.org/abs/2301.06011}{{\ttfamily arXiv:2301.06011
  [hep-th]}}.

\bibitem{Henneaux:2021yzg}
M.~Henneaux and P.~Salgado-Rebolledo, ``{Carroll contractions of
  Lorentz-invariant theories},''
  \href{http://dx.doi.org/10.1007/JHEP11(2021)180}{{\em JHEP} {\bfseries 11}
  (2021) 180}, \href{http://arxiv.org/abs/2109.06708}{{\ttfamily
  arXiv:2109.06708 [hep-th]}}.

\bibitem{Henneaux:1979vn}
M.~Henneaux, ``{Geometry of Zero Signature Space-times},'' {\em Bull. Soc.
  Math. Belg.} {\bfseries 31} (1979) 47--63.

\bibitem{Bergshoeff:2017btm}
E.~Bergshoeff, J.~Gomis, B.~Rollier, J.~Rosseel, and T.~ter Veldhuis,
  ``{Carroll versus Galilei Gravity},''
  \href{http://dx.doi.org/10.1007/JHEP03(2017)165}{{\em JHEP} {\bfseries 03}
  (2017) 165}, \href{http://arxiv.org/abs/1701.06156}{{\ttfamily
  arXiv:1701.06156 [hep-th]}}.

\bibitem{Campoleoni:2022ebj}
A.~Campoleoni, M.~Henneaux, S.~Pekar, A.~P\'erez, and P.~Salgado-Rebolledo,
  ``{Magnetic Carrollian gravity from the Carroll algebra},''
  \href{http://dx.doi.org/10.1007/JHEP09(2022)127}{{\em JHEP} {\bfseries 09}
  (2022) 127}, \href{http://arxiv.org/abs/2207.14167}{{\ttfamily
  arXiv:2207.14167 [hep-th]}}.

\bibitem{Horndeski:1974wa}
G.~W. Horndeski, ``{Second-order scalar-tensor field equations in a
  four-dimensional space},'' \href{http://dx.doi.org/10.1007/BF01807638}{{\em
  Int. J. Theor. Phys.} {\bfseries 10} (1974) 363--384}.

\end{thebibliography}\endgroup

		\end{document}